\documentclass[12pt,preprint]{aastex}
\usepackage{natbib}
\usepackage{amsmath}

\newcommand{\el}{$\ell$}
\newcommand{\Vosc}{$V_{osc}$}
\newcommand{\Pdot}{\textit{\.{P}}}
\newcommand{\dPdot}{\textit{$\Delta$\.{P}}}
\newcommand{\dPdotcrit}{\textit{$\Delta$\.{P$_{crit}$}}}
\newcommand{\dVosc}{$\Delta{V_{osc}}$}
\newcommand{\del}{$\Delta\ell$}
\newcommand{\gmodes}{{\it g-}modes}
\newcommand{\gmode}{{\it g-}mode}
\newcommand{\pmodes}{{\it p-}modes}

\begin{document}

\title{A Non-Radial Oscillation Model for Pulsar State Switching}

\author{R. Rosen}
\affil{NRAO, P.O. Box 2, Green Bank, WV 24944}
\email{rrosen@nrao.edu}
\author{M. A. McLaughlin}
\affil{West Virginia University, 436 Hodges Hall, Morgantown WV 26506, Also adjunct at the National Radio Astronomy Observatory}
\author{S. E. Thompson}
\affil{SETI Institute/NASA Ames Research Center, M/S 244-30, Moffett Field, CA 94035, USA}

\begin{abstract} 
Pulsars are unique astrophysical laboratories  because of their clock-like timing precision, providing new ways to test general relativity and detect gravitational waves. One impediment to high-precision pulsar timing experiments is timing noise.  Recently \citet{lyn10} showed that the timing noise in a number of pulsars is due to quasi-periodic fluctuations in the pulsars' spin-down rates and that some of the pulsars have associated changes in pulse profile shapes. Here we show that a non-radial oscillation model based on asteroseismological theory can  explain these quasi-periodic fluctuations. Application of this model to neutron stars will increase our knowledge of neutron star emission and neutron star interiors and may improve pulsar timing precision.
\end{abstract}

\section{Introduction}

Pulsars have long been known as superb laboratories for fundamental physics due to their timing precision and stable pulse profiles. Timing observations of pulsars have resulted in the best tests of general relativity in the strong-field regime, the most precise stellar mass measurements, and the most constraining limits on gravitational wave sources in the nanoHertz regime through `pulsar timing array' experiments \citep[e.g.][]{jen06,kra06a}. One significant limitation for all of these studies is timing noise, or non-random variations in the residuals between measured and model-predicted arrival times.  \citet{lyn10} recently showed that the timing noise in 17 pulsars is due to the pulsar's spin-down rate fluctuating between different values in a quasi-periodic manner. Furthermore, for six of these pulsars,  the changes in the spin-down rate directly correlate to changes in the pulse shape.  While the pulsars described in \citet{lyn10} have long periods, some millisecond pulsars, such as PSR B1931+27, also exhibit timing noise and may show similar but less obvious effects. This behavior offers a possible prescription for removing timing noise from data and dramatically increasing timing stability, facilitating projects such as the direct detection of gravitational waves through pulsar timing.  In this paper, we present a model for this behavior based on non-radial oscillations.  We outline the model in \S\ref{model}, discuss underlying assumptions in \S\ref{theory}, describe the implications for the pulsars discussed in \citet{lyn10} in \S\ref{predictions}, and present conclusions in \S\ref{conc}.

\section{A Non-Radial Oscillation Model}
\label{model}

\begin{figure}
\includegraphics[scale=0.6]{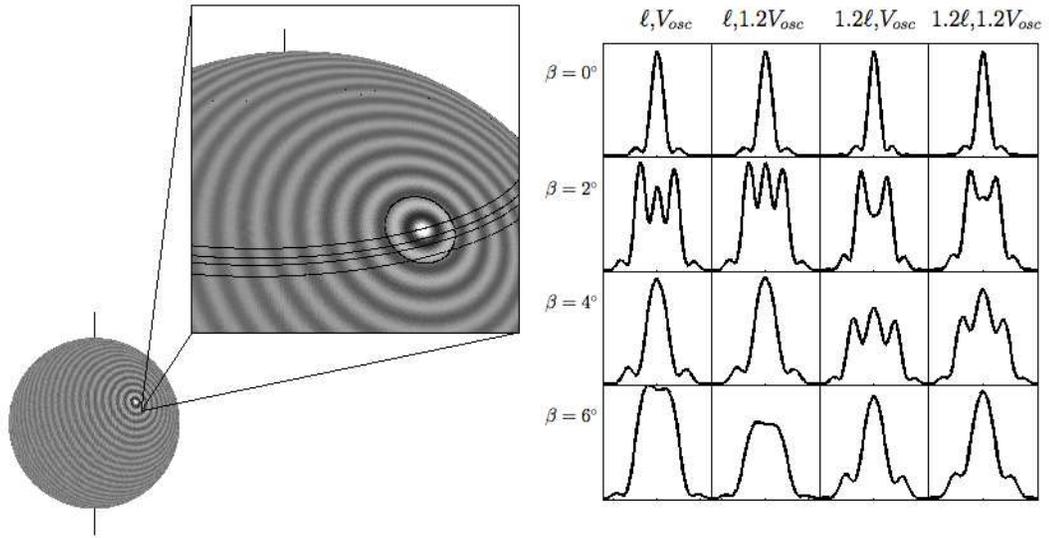}
\caption{\textit{Schematic}: A map of the non-radial oscillations on a pulsar.  The circle around the magnetic pole in the enlarged view denotes the boundary of the emitting region.  Dark regions indicate negative displacements and light regions positive ones. After a half cycle of the oscillations, the dark regions would be light and vice versa, but the nodal lines (as defined by \el) separating them would remain unchanged, except for rotation of the whole pattern about the rotation axis of the star. The magnetic cap is crossed by four representative sightlines, where $\beta$ is the angle between our sightline and the magnetic cap.  The listed values for $\beta$ are equally spaced.  \textit{Panels}: For each sightline, the columns show the corresponding average profile (i.e. intensity vs. pulse phase) for various values of \Vosc~ and \el.}
\label{fig1}
\end{figure}

We show that a non-radial oscillation model can explain all the long-period variations described in \citet{lyn10}. In our model, based on astroseismological theory \citep{dzi77}, non-radial oscillations of high spherical degree (\el) are aligned to the magnetic axis of the neutron star \citep{cle04}. Between any two adjacent nodal lines, displacements of material move horizontally along the stellar surface toward and away from the magnetic pole, as shown in the schematic of Figure \ref{fig1}.   \citet{jes01} showed that thermal emission of electrons dominates the rate at which charged particles flow from the surface into the magnetosphere.  Non-radial oscillations, and their associated velocities, will vary the amount of thermal emission by creating regions of local heating due to compression between the nodal lines, thus increasing the flow of charged particles into the magnetosphere and modulating the radio emission \citep{str92}.  
 
\begin{figure}
\includegraphics[scale=0.5]{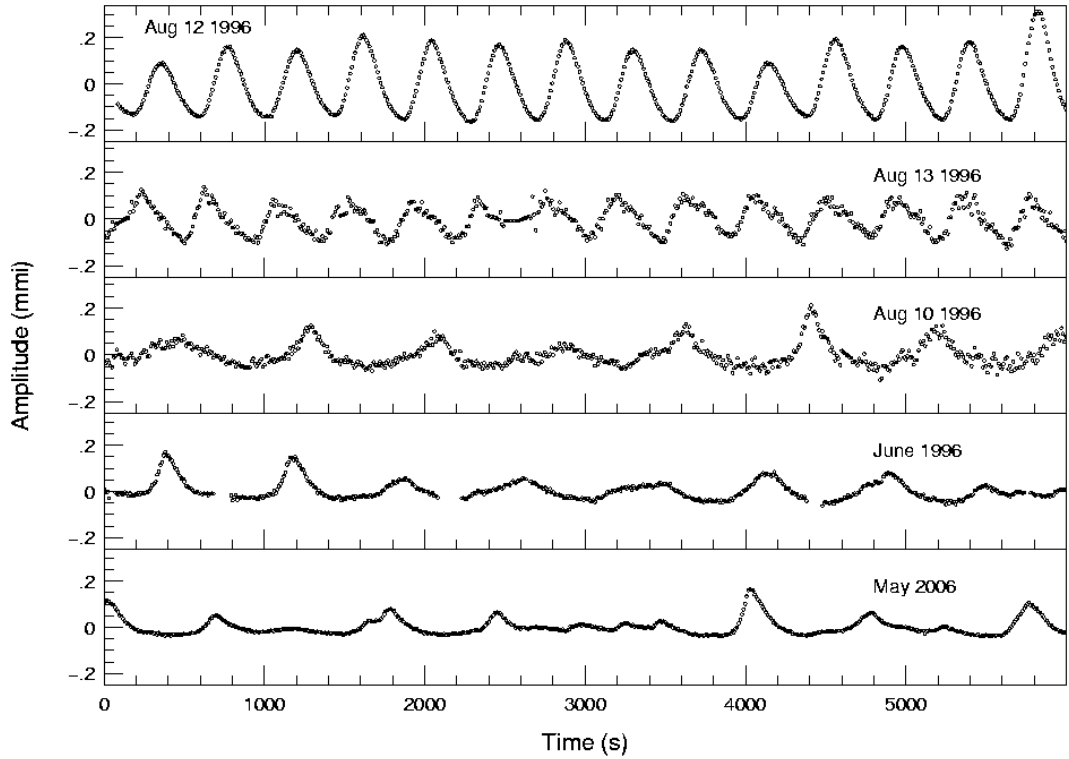}
\caption{The flux variations of GD358, a helium atmosphere pulsating white dwarf star, observed on different nights \citep{prov09}. For two nights in 1996, the light curve of GD358 drastically changed. Unlike during all other observations of this star, all the pulsation energy was pumped into the shortest period modes. Reproduced by permission of the AAS}
\label{fig2}
\end{figure}

The situation is analogous to behavior seen in white dwarf stars.  Some white dwarfs are \gmode, multi-periodic pulsators that tend to excite a select number of pulsation modes, as shown in Figure \ref{fig2}.  In adiabatic oscillations, local heating of the stellar surface coincides with pulsation maximum during the compression phase of the pulsation.  A quarter cycle later, material will be moving away at a maximum velocity.  The displacements responsible for surface heating and flux changes in oscillating white dwarfs are primarily non-radial because of their high surface gravity \citep{rob82} and have measured radial velocity variations that can be interpreted as the horizontal velocities \citep{vank00,tho03}.  In neutron stars, we expect that the amplitude of the displacements follows surface thermal variations caused by non-radial oscillations of the neutron star.  As with the white dwarf stars, an increase in the velocities can result in increased thermal emission from the star.  

In our non-radial oscillation model, each frequency of oscillation has its own periodic velocity (or displacement) amplitude.  In the pulsars we have studied to date, the amplitude of the displacements dictates the average pulse shape \citep{ros08,ros10}, but as the amplitude of the velocities grows, the average pulse shape can change \citep{cle08}.   In these pulsars, through fitting a non-radial oscillation model to their drifting subpulses, we have also found \el~ to be large compared to that seen in white dwarf stars, consistent with their smaller size.  For pulsars B0809+74 and  B0943+10, \el~ values range from $126 \leq$ \el~ $\leq 133$ \citep{ros10} and $385 \leq$ \el~ $\leq 875$ \citep{ros08}, respectively.  In white dwarf stars, modes with high \el~ fade from view because of geometric cancellation of the stellar surface, leaving only modes of low \el~ (\el~ $\leq 4$) \citep{yea05,tho04,tho08}.  

By fitting a non-radial oscillation model to the drifting subpulses in PSRs B0943+10 and B0809+74, we measured the pulsation period to be on the order of 30$-$50 ms \citep{ros08,ros10}.  These values for the pulsation period are consistent with core \gmodes~ \citep{rei92}.  However, core \gmodes~ require large excitation energies, and therefore surface \gmodes~ are a possibility \citep{str93} as they have lower energies and larger surface amplitudes than the core modes \citep{mcd88}.  The period predictions for core \gmodes~ in neutron stars range from a minimum value of 10 ms \citep{rei92} to 2$-$88 seconds \citep{mcd88} depending on the model for the structure and composition of the stellar interior.  Surface \gmodes~ have periods in the 40$-$400 ms range \citep{mcd88}.  While the pulsation periods calculated for most models are calculated assuming low \el, the period of the mode decreases for higher spherical degree \citep{mcd88}.  It is unlikely that the pulsation modes we see in PSRs B0943+10 and B0809+74 are \pmodes~ as these modes have periods on the order of tenths of milliseconds \citep{mcd88} and have overtones with shorter periods, whereas the overtones of \gmodes~ have longer periods.

The two oscillation driving mechanisms for white dwarf stars are the $\kappa-\gamma$ mechanism and convective driving.  For pulsating DOVs stars like PG1159$-$035, characterized by an atmosphere that lacks hydrogen but shows helium, carbon, and oxygen absorption lines, the $\kappa-\gamma$ mechanism drives the pulsations as the opacity varies steeply with pressure \citep{sta83,cor06,cox03}.  In DAV and DBV stars, which have atmospheres dominated by hydrogen or helium, respectively, convective motions drive pulsations \citep{bri91,gol99,wu99}.

Driving mechanisms in for neutron star oscillations are different and less well understood.  \citet{jes01} show that thermal and field emission from the neutron star surface can accelerate electrons along open field lines with the formation of a vacuum gap.  Therefore, oscillations should be related to thermal variations at the neutron star surface, as would be produced by \gmodes.  The high spherical degree that we observe in the data (and is predicted by our model) suggests that the pulsation driving energy is concentrated in a small area on the surface of the star.  Otherwise it would average away over multiple surface zones \citep{cle04}.  In one possible scenario, the source of the driving energy is at the magnetic cap and the particle emission exerts a torque on the open field lines, which is coupled to the rest of the star via magnetic and mechanical dissipation.  If the magnetic field is coupled (even weakly) to the surface and it displaces material laterally, the torque on the open field lines results in a displacement of material on the polar cap. The heating that results can increase particle emission, which then increases the torque, and drives the oscillations \citep{cle04}.  

Other possibilities for driving mechanisms involving direct shaking of field lines \citep{bor76}, thermal stratification \citep{mcd83}, chemical stratification \citep{fin87}, and stratification in the core \citep{rei92}.  The gravitational radiation damping timescale for \gmodes~ exceeds the age of the Universe and is therefore a negligible effect \citep{mcd83,rei92}.  The pulsations we observe in PSRs B0943+10 and B0809+74 appear to persist on hour to month (if not longer) timescales and any driving mechanism must  be in agreement with these observations.  Other neutron star oscillations have been detected in the form of quasi-periodic oscillations (QPOs) from soft gamma-ray repeaters.  These have been interpreted as torsional shear modes; the frequencies of QPOs are too high to originate from \gmodes~ \citep{wat07}. 

\section{Standard Model Assumptions}
\label{theory}
 
In the standard pulsar model \citep{gj69}, the magnetosphere is considered to be `force-free' in that the electromagnetic energy density dominates all other dissipative forces and that radiative emission has a negligible effect on the spin-down rate \citep{aro07}.  Instead, the configuration of the magnetosphere, based on the polar cap cascade zone and the current density within the open field lines, and its change over time dictates the pulsar spin-down rate \citep{tim06,aro07,tim10}.  Recently, \citet{tim10} speculates that the magnetospheres of some pulsars can have several quasi-stable states with different configurations and that the switching of the magnetosphere between these states can result in the observed mode changes and nulls.  \citet{tim10} further notes that each state would have a different spin-down rate.

\citet{tim07} shows that oscillation modes of high spherical degree and non-vanishing velocities at the surface can alter the local Goldreich-Julian (GJ) charge density  \citep{gj69}.  The GJ charge density, while not a contributing factor to the spin-down rate in the force-free model, is then a combination of the charge density due to rotation and the charge density due to oscillations.  In this paper, we assume that, like white dwarf stars, neutron star oscillations can change with both amplitude and spherical degree, therefore affecting the total charge density and the magnetospheric configuration.  If the charge density is at least partially responsible for the energy loss of the pulsar and/or the change in pulsation mode changes the magnetospheric configuration, we expect to observe a change in spin-down rate.

\section{Theoretical Predictions}
\label{predictions}

Under some models we expect for the spin-down rate and the pulse shape to change with both spherical degree and the amplitude of the velocities of the oscillations, \Vosc.  \citet{tim07} calculates the GJ charge density for neutron star oscillations assuming the standard pulsar model of a rotating magnetized conducting sphere surrounded by plasma, and explores the excitation of oscillations by neutron star glitches.  He shows that oscillation modes of high spherical degree and non-vanishing velocities at the surface can alter the local GJ charge density ($\rho_{GJ}$) where $\rho_{GJ} \sim$ \el$(V_{osc}/c)(B/4\pi{R_{NS}})$.  The charge density will affect the accelerating electric field.  For the oscillations to be strong enough to influence the particle distribution, resulting in effects that we can observe, \citet{tim07} suggests the spherical degree is approximately several hundred, which is in agreement with our independent measurements \citep{ros08,ros10}.  Because more particles leave the stellar surface when the charge density is greater, we expect the spin-down rate to increase with either: 1) increasing \el, 2) increasing \Vosc, or 3) increasing \el~ and \Vosc.  Assuming that the current density in the magnetosphere is given by Equation \ref{eqn:1} \citep{hbpa}, the charge density is directly related to the spin-down rate by:

\begin{equation}
\frac{\rho_{GJ}}{e} = -\frac{{\Omega{B}}}{2{\pi}ce} \simeq 7 \times 10^{10} \mathrm{cm}^{-3} \left(\frac{P}{s}\right)^{-1/2}\left(\frac{\dot{P}}{10^{-15}}\right)^{1/2}
\label{eqn:1}
\end{equation}
\\where ${\Omega}$ is the angular rotation frequency, ${B}$ is the magnetic field, $P$ is the spin period of the pulsar, and \Pdot~ is the spin-down rate.  Note that this assumption is not in agreement with the simple force-free model.  We calculate the fractional change in charge density using the change in spin-down rate and then use this to calculate $\rho_2/\rho_1 =$ \el$_2V_{osc,2}$/\el$_1V_{osc,1}$.  For the pulsar with the largest change in spin-down rate, PSR B1931+24, the value of  \el\Vosc~  will change by 20\%; for PSR B1929+10, which has the smallest change in spin-down rate, \el\Vosc~  will change by 0.15\%.  To make these estimates, we assume that the contribution to the charge density due to oscillations is much greater than the contribution due to rotation.  Figure \ref{fig1} illustrates how changes in these parameters affect the pulse shapes and Table \ref{tab1} shows our best guesses for the classifications of the pulsars in \citet{lyn10} into the different scenarios.
 
In the first scenario where \Vosc~ changes but \el~ remains constant, pulsars fall into two categories:  small and large velocity oscillations.  For small velocity oscillations, the change in the spin-down rate is less than some value, \dPdotcrit, and is accompanied by little or no change in the pulse shape.  With the small sample of pulsars that show this behavior in \citet{lyn10}, we cannot determine the precise demarcation point between large and small changes in the spin-down rate, but estimate it to be \dPdotcrit~ $\sim$10\% by inspection of those pulsars which show pulsar shape changes.  The pulsars in the first column of Table \ref{tab1} fall into this category.  Many stable pulsating white dwarf stars have subtle changes in amplitude over time, thereby changing \Vosc~ but not \el~ \citep{prov09,vui00,han08}.  When the amplitude of a single mode changes, both its flux and velocity amplitudes change.  If amplitude changes occur on a neutron star, the change in \Vosc~ will also change the spin-down rate.  A periodic change in amplitude will result in a periodic change in the spin-down rate.  Because the spherical harmonic is responsible for the average pulse shape and \el~ is constant, the average pulse shows little or no change with varying spin-down rate.

\begin{table}
\begin{tabular}{cccc}
Scenario 1 & Scenario 1 & Scenario 2 & Scenario 3 \\
\hline
\dPdot $<$ \dPdotcrit & \dPdot $>$ \dPdotcrit & \dPdot $<$ \dPdotcrit & \dPdot $>$ \dPdotcrit \\
no pulse shape changes & small pulse shape changes & pulse shape changes & large pulse shape changes\\
small \dVosc, no \del & large \dVosc, no \del & no \dVosc, \del & \dVosc, \del \\
\hline
B1642$-$03 & B2035+36 & J2043+2740 & B1931+24\\
B1903+07 & ...  & B1822$-$09 & ...\\
B1839+09 & ... & B1540$-$06 & ... \\
B2148+63 & ...& B1828$-$11 & ...\\
B1818$-$04 & ... & B0740$-$28 & ...\\
B0950+08 & ... & ... & ...\\
B1714$-$34 & ... & ... & ...\\
B1907+00 & ... & ... & ...\\
B1826$-$17 & ... & ... & ...\\
B0919+06 & ... & ... & ...\\
B1929+20 & ... & ... & ...\\
\end{tabular}
\caption{List of pulsars which show changes in the spin-down rate with possible classifications with respect to non-radial oscillation properties.}
\label{tab1}
\end{table}

Pulsars with large velocity oscillations have changes in the spin-down rates greater than \dPdotcrit~ with changes in the pulse shape, such as B2035+36, are listed in the second column of Figure \ref{fig1}.   In the small sample of pulsars we have quantitatively fit, the average pulse shape is dominated by the amplitude of the displacements \citep{cle08}.  However, when the relative amplitudes of the velocities and displacements becomes dominated by the velocities, the pulse shape can change and we may observe a corresponding change in the polarization behavior as well \citep{cle08}.  PSR B2035+36 shows a sudden large change in spin-down rate.  Analogously, the white dwarf star GD358 suddenly pumped all of its energy into the shortest period modes, increasing the overall amplitude and changing the pulsation characteristics on a timescale of hours as shown in Figure \ref{fig2} \citep{kep03,prov09}.  Such exchanges in the energy between pulsations on white dwarfs is not always so dramatic but are common \citep{kle98,vui00,wu01}.

Pulsars that have changes in the spin-down rate less than \dPdotcrit~ accompanied by changes in the pulse shape, like those listed in column three in Table 1, fall into the second scenario, where \el~ varies but \Vosc~ remains constant, as illustrated in the third column of Figure \ref{fig1}.  Periodic changes in \el~ will modulate both the spin-down rate, the pulsation period, and the average pulse profile.  White dwarf stars are known to variably excite pulsations of \el, varying from $1\leq$ \el~$\leq 4$ \citep{kep03, tho08, cle00}.  When oscillations of different \el~ are excited on a pulsar, the relationship between the nodal structure to the pulse width is more complicated than a simple proportion due to the geometry of the star and our viewing angle relative to the magnetic pole.  While it is easy to assume that as the area between nodal lines gets larger the pulse width increases; for certain geometries, the pulse width can actually decrease, as in PSR J2043+2740 and as illustrated in Figure \ref{fig1}.  Radio polarization data would allow us to determine the geometry and the amount by which \el~ is changing. 

The third scenario, where both \Vosc~ and \el~ change, can also be possible through the transfer of energy between modes.  This is shown in the fourth column of Figure \ref{fig1} and Table \ref{tab1}.  If both \Vosc~ and \el~ vary, we expect larger changes in the spin-down rate and pulse width, like that seen in PSR B1931+24.  If \Vosc~ and \el~ are changing at different rates, we also expect to see more complicated behavior including beat frequencies which would result in complicated patterns in the spin-down rate and pulse shape.  Extrapolating the more complex behavior of unstable white dwarf pulsators to neutron stars, we would expect to see multiple, possibly simultaneous pulsation modes.  The changes in the light curves in white dwarf stars are not necessary gradual or sinusoidal and therefore we do not necessarily expect gradual changes in the spin-down and pulse shape in neutron stars.  The velocity and spherical degree could change depending on which mode is dominating the pulsation spectrum.  


All the scenarios above can explain how the spin-down rate can change periodically with or without accompanying changes in the pulse profile.  It may be that some pulsars do not exhibit any changes in the spin-down rate with time but still show pulsations.  These objects would have stable pulsation modes.  Stable white dwarf pulsators can have several pulsation modes present simultaneously and tend to be dominated by a few pulsations \citep{yea05, mul08}.  We speculate that neutron stars which do not exhibit changes in spin-down rate or period show no changes in \Vosc~ or \el, but may have several pulsations modes present.  PSRs B0943+10 and B0809+74 have both been quantitatively fit to an oscillation model but do not exhibit any known changes in the spin-down rate \citep{ros08,ros10}.  Just as only a percentage of white dwarf stars exhibit pulsations, we expect only a fraction of neutron stars to show oscillations.

\section{Conclusion}
\label{conc}

In summary, we have shown that the fluctuations in spin-down rate described in \citet{lyn10} can be explained by a non-radial oscillation model.  Our model is not complete, as little is understood about pulsar magnetospheres and there is no clear mechanism for driving the oscillations.  However, our oscillation model does not require invoking an external influence to explain the periodicity, as in \citet{cor08}, and is based on observations of other well studied stars.  This will lead to better understanding of pulsar emission physics and may enable optimal removal of  effects due to non-radial oscillations in pulsar timing data, increasing the usefulness of pulsars as fundamental physics laboratories.

\acknowledgements
MAM is supported by WVEPSCOR, the Research Corporation, the Sloan Foundation, and the National Science Foundation.  The National Radio Astronomy Observatory is a facility of the National Science Foundation operated under cooperative agreement by Associated Universities, Inc.  Also we thank Duncan Lorimer and Xavier Seimens for helpful comments on the manuscript.

\end{document}